\documentclass[a4paper]{article}
\usepackage{icrc2013}
\pdfoutput=1

\title{Status of the new Sum-Trigger system for the MAGIC telescopes}

\shorttitle{MAGIC: Sum-Trigger-II}

\authors
{
	J.R. Garc\'{\i}a$^{*,1,3}$,
	F. Dazzi$^{2}$, 
	D. H\"{a}fner$^{1}$, 
	D.Herranz$^{4}$, 
	M. L\'opez$^{4}$, 
	M. Mariotti$^{2}$, 
	R. Mirzoyan$^{1}$,  
	D. Nakajima$^{1}$, 
	T. Schweizer$^{1}$, 
	M. Teshima$^{1}$.
}

\afiliations
{
	$^1$ Max Planck Institut f\"{u}r Physik.\\
	$^2$ University of Padova, INFN Sezione di Padova.\\
	$^3$ Instituto Astrofisico de Canarias.\\
	$^4$ Universidad Complutense de Madrid.\\
	$^{*}$ Presenting author
}
\email{jezabel@mpp.mpg.de}

\abstract{MAGIC is a stereoscopic system of two 17\,m diameter Imaging Air Cherenckov Telescopes (IACTs) for $\gamma$-ray astronomy. Lowering the energy threshold of IACTs is crucial for the observation of Pulsars, high redshift AGNs and GRBs. A lower threshold compared to conventional digital trigger can be achieved by means of a novel concept, the so called Sum-Trigger, based on the analogue sum of a patch of pixels. The Sum-Trigger principle has been proven experimentally in 2007 by decreasing the energy threshold of the first MAGIC telescope from 55\,GeV down to 25\,GeV. The first VHE detection of the Crab Pulsar was achieved due to this low threshold. After the upgrade of the MAGIC I and MAGIC II, a new Sum-Trigger system will be installed in both telescopes in Summer 2013. The expected trigger threshold in stereo mode is about 25\,$\div$\,30\,GeV. It is a an improvement over the existing threshold (about 50\,GeV) of the digital trigger. We will report about the current status of the project.}

\keywords{Magic, Hardware, trigger, analogue, Sum-Trigger, Crab.}

\begin{document}
\maketitle

%Begin a section.
\section{Introduction}
One of the main attempt of $\gamma$-ray Astrophysics detector development is to cover the energy range from a few GeV to 80\,GeV, since this region is difficult to access, for both, space-based and ground-based instruments. There are strong scientific motivations to extend the IACT technique below 80\,GeV. 
The main goal is to study transient objects such as high redshift AGNs and GRBs, which are not easily detectable by satellites due to the low flux. In addition, lowering the threshold also favourites both the study of EBL and the measurements of the spectra of VHE Pulsars \cite{bib:pulsar}. 

Therefore, the development of new strategies to decrease the trigger thresholds and increase the sensitivity in this gap is crucial. In 2008, MAGIC already succeeded in this sense by developing a prototype analogue Sum-Trigger (single telescope) that significantly lowered the trigger threshold down to 25\,GeV. 
This trigger system achieved excellent results, since it allowed the first detection of very high energy pulsed gamma radiation from the Crab Pulsar \cite{bib:pulsar}.

Using the experience acquired during the construction and operation of the first prototype, a new analogue trigger, so-called Sum-Trigger-II, was designed and developed. This paper aims to explain the concept, development, and status of this system.

\subsection{Description of the MAGIC experiment}
The MAGIC telescopes are located at a height of 2200\,m a.s.l. on the Observatorio del Roque de los Muchachos (28\hspace{-1.5mm}$\phantom{a}^{\circ}$N, 
18\hspace{-1.5mm}$\phantom{a}^{\circ}$W) on Canary Islands. It is a stereo system composed of two 17\,m diameter (f$/$D $\sim$1) IACTs for VHE $\gamma$-rays observation. Both telescopes are built using the same light-weight carbon-fibre structure, which allows for a rapid repositioning time necessary for observations of short phenomena, such as GRBs. The camera of each telescope is composed of 1039, 0.1\hspace{-1.5mm}$\phantom{a}^{\circ}$ hexagonal pixels (PMTs), it has a 3.5\hspace{-1.5mm}$\phantom{a}^{\circ}$ field of view and a trigger area of 1.25\hspace{-1.5mm}$\phantom{a}^{\circ}$ radius. In each pixel the signals from the PMTs are optically transmitted to the counting house where the trigger and the digitization of the signals take place.

Since 3 years, regular observations are performed in stereoscopic mode using the standard trigger system, composed of three stages. The level-zero trigger sets the discriminator threshold in real time for each pixel in the trigger region, excluding the electronic noise and part of the Night Sky Background (NSB). Each telescope separately has a level-one digital trigger with the 3 next neighbour (3NN) topology, it means that it only selects close compact events that include at least 3 next neighbouring pixels. Only events that trigger both telescopes are recorded. The stereo trigger, level-three, makes a tight time coincidence between both telescopes. 

Aiming for the lowest possible threshold, as an alternative to the current standard trigger (level-zero plus one), the Sum-Trigger-II will be installed by end of this Summer. 

\subsection{Low energy observations with IACTs}
The physics principles of cascades in extended air showers are the same in all the energy ranges, but the observation of low energy cascades is more difficult due to the fact that less energy gets converted while the particles are crossing the atmosphere. The cascades reach their maximum development around a depth of 180-240\,gr/cm$^2$ (11\,$\div$\,12\,km a.s.l.), very far away from the position of the IACTs thus a large part of the shower photon signal is absorbed in the
atmosphere. Adding the fact that the energy threshold for the Cherenkov photon production is higher in the upper layers of atmosphere due to reduced air density
(35\,$\div$\,40\,MeV for electrons), the number of emitted photons is further reduced. On top, the higher is the altitude, the lower is the refraction index and thus, the smaller is the Cherenkov angle, and the generated photons are strongly collimated along the charged particles trajectory. In the case of showers initiated by $\gamma$-rays, which have a modest transversal development, the main consequences are small shower projections (small size) and a limited number of triggered events at large impact parameters.

\begin{figure}[h]
	\centering
  	\includegraphics[width=0.45\textwidth]{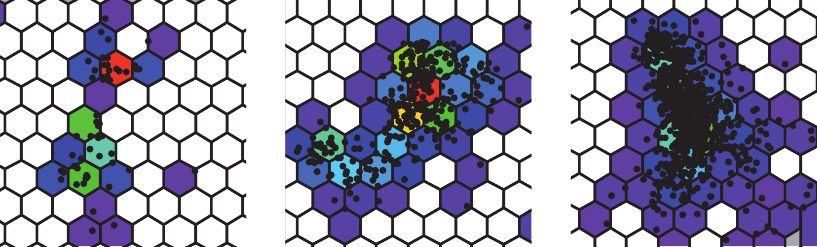}
  	\caption{Distribution of photons (black points) at camera level for three different $\gamma$-ray energies, from left to right: 12.9\,GeV, 21.5\,GeV $\&$ 58.6\,GeV. Adapted from \cite{bib:Dazzi}.}
  	\label{low_fig}
\end{figure}

Monte Carlo simulations level of the morphology of low energy events ($\sim$10\,$\div$\,60\,GeV) have been done at the camera \cite{bib:Dazzi}. These studies revel that most of these events are concentrated in a thin doughnut shaped region around the camera center. Although the primary shower energy is very low, the photons span on a wide region. Fig.\ref{low_fig} shows three typical gamma showers between about 13\,GeV and 60\,GeV. However, the photon density in most of the single pixels is very low ($\le$ 10 photons, violet and blue pixels) and only few of them contain enough signal to be detected (green, yellow and red). Usually, up to 25\,GeV, 2 or 3 groups or islands of signals are observed. At higher energies these islands start to
overlap and the ovoid like image of the Cherenkov light is recovered. 

All these studies led to the construction of a new trigger system, the Sum-Trigger, particularly sensitive to low energy cascades.

\section{The Sum-Trigger-II}
\subsection{The concept}
\begin{figure}[h!]
	\centering
  	\includegraphics[width=0.38\textwidth]{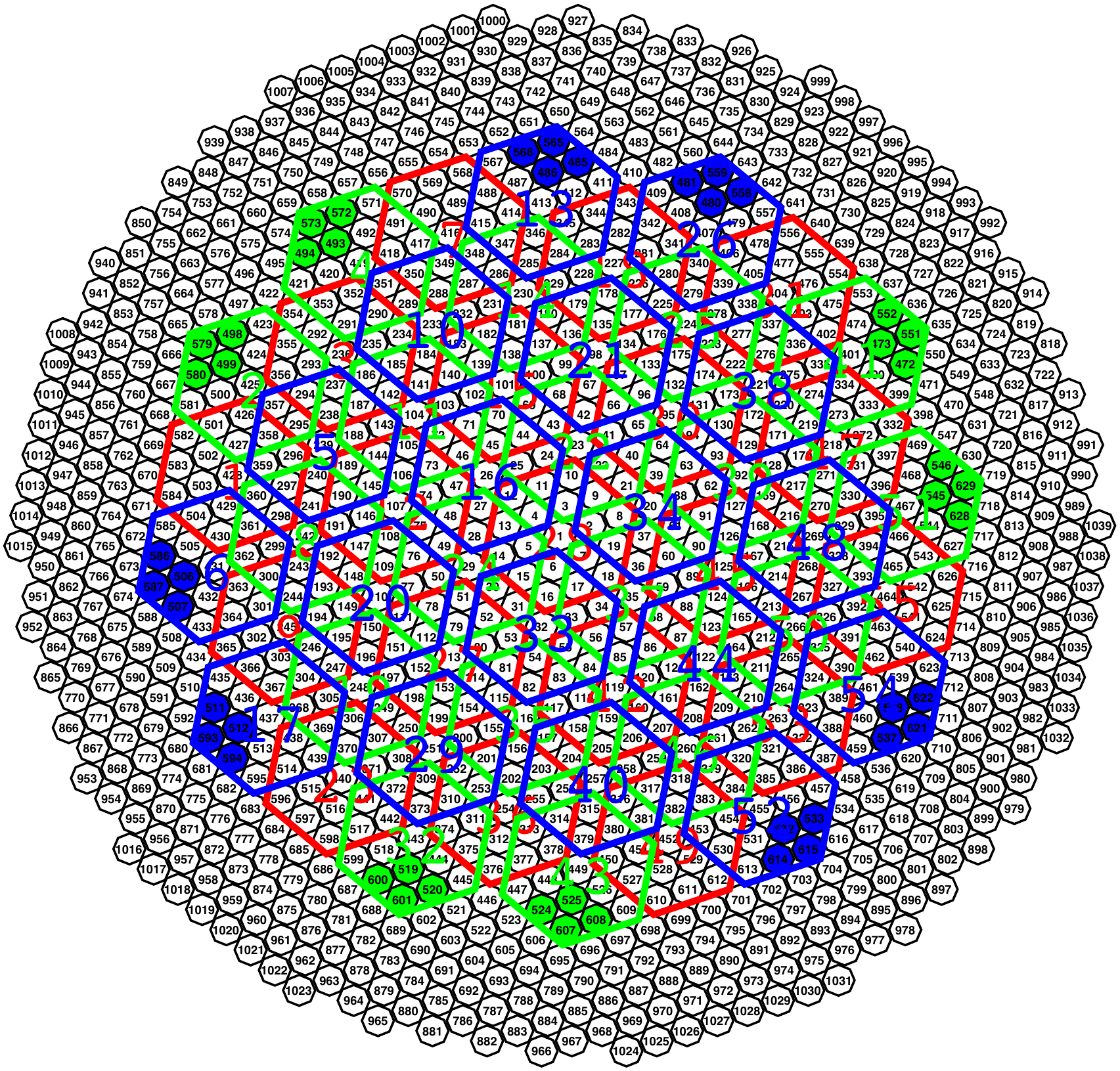}
  	\caption{Layout of the trigger region of Sum-Trigger-II, where each color represents a different layer. 19 pixel hexagonal macrocell shape was selected to guarantee both a symmetrical overlap and an angular symmetry.}
  	\label{macro_fig}
\end{figure}

The basic principle of the Sum-Trigger-II is to add up the signals from a group of neighbouring pixels (macrocell) and then apply a threshold to the summed signal. 
For this purpose the trigger region has been covered with three overlapping layers of macrocells, see Fig.\ref{macro_fig}. Monte Carlo studies suggested that the optimum solution is to use roundish macrocells composed by 19 pixels (of size 0.1 deg), then a macrocell contains most part of a low energy shower image.

One of the advantages of this technique is that it improves the signal to noise ratio (shower signal to NSB), since all pixels within a macrocell contribute to the trigger decision, i.e. it uses small photon signals below the single channel threshold (below which the standard trigger is sensitive to).

Besides the simplicity of the idea, the Sum-Trigger-II construction has required the development of fast, high precision, analogue electronics. The PMTs in the cameras do not have exactly identical properties, but there are differences in signal propagation times, gains and pulse widths. \\
In addition, these different signal delays will be worsened due to small length differences in the 162\,m long optical fibres that transport the signals from the camera to the readout electronics. The skew must be minimized in the whole trigger area to assure a proper pile up among signals.

The pulse width itself is perfectly optimized for the temporal evolution of a shower core started from a low energy gamma event. If it is too small then the signal of the shower does not sum up at the same time, if it is too slow then one integrates too much night sky noise. The optimal pulse's width was determined to be between 2.5\,ns and 3.0\,ns FWHM.

The best threshold level for a patch of 19 pixels with an acceptable ratio between accidental NSB and lowest trigger energy threshold is around 24\,$\div$\,27\,PhE.

A drawback of classical PMTs is so-called afterpulse noise, which are random noise pulses of high amplitude and high rate. Such afterpulses could trigger the Sum-Trigger-II and need to be vetoed. The simple way of reducing the effect of those afterpulses is to clamp the signal at a certain (optimized) amplitude. For the PMTs that are used in MAGIC the optimum is around 6\,$\div$\,8\,PhE clipping level for the single telescope (corresponding to avoiding triggers in
a three-fold afterpulse coincidence). In stereo mode (coincidence between two telescopes) this clipping level can be relaxed (increased) which further improves the sensitivity at lowest energies.

\begin{figure}[h]
	\centering
	\includegraphics[width=0.45\textwidth]{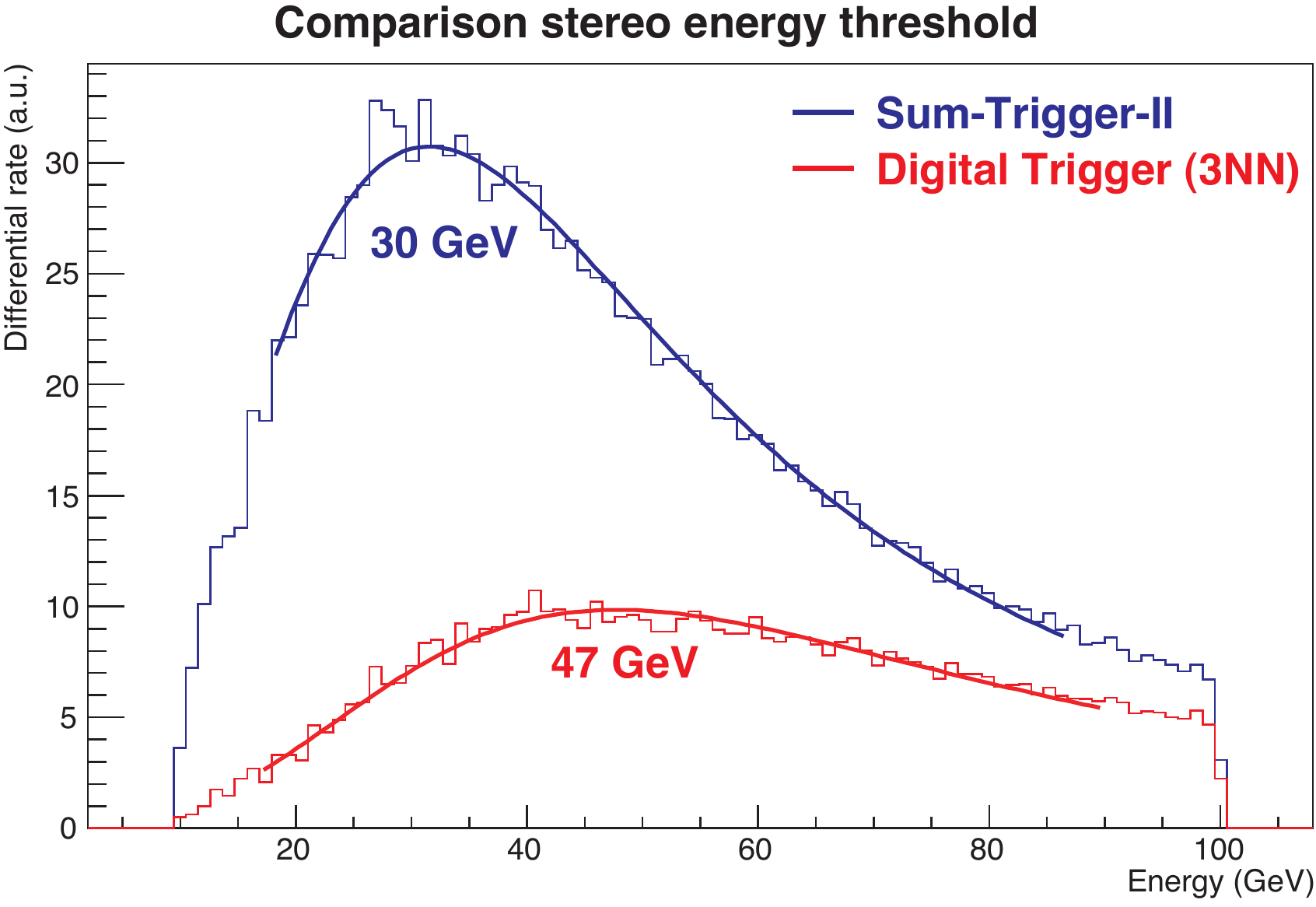}
	\caption{Monte Carlo simulations show the threshold comparison between the digital trigger and the Sum-Trigger-II.} 
	\label{Monte-Carlo_sum-trigger}
\end{figure}

The Monte Carlo studies (Fig.\ref{Monte-Carlo_sum-trigger}) show the improvement on the sensitivity and the lower threshold (around 30\,GeV) respect to the 3NN digital trigger already stated \cite{bib:Dazzi}.

\newpage
\subsection{Functional set-up of the overall trigger system}
\begin{figure}[ht]
	\centering
	\includegraphics[width=0.45\textwidth]{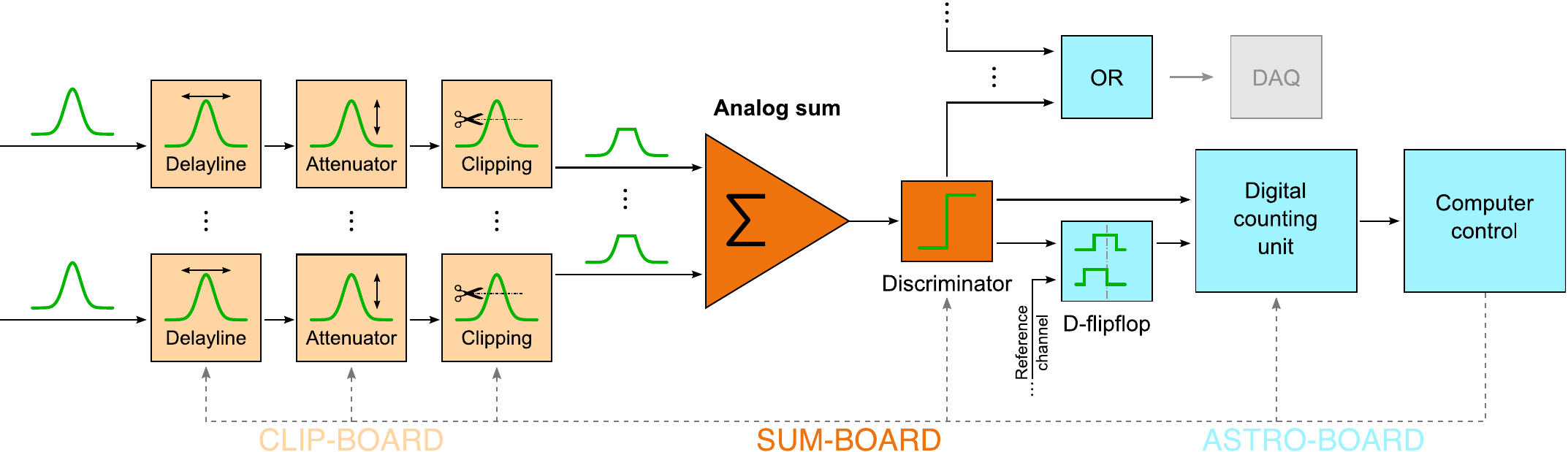}
	\caption{ Sum-Trigger-II schematics. Each channel has an analogue delay adjustment, an amplitude adjustment and a clipping mechanism. The sum of 19 channels is fed into a discriminator.} 
	\label{ST_schematics_fig}
\end{figure}

The Sum-Trigger-II system is structured in several subsystems (see Fig.\ref{ST_schematics_fig}). The signals first arrive to so-called Clip-boards. After delaying, adjusting amplitude and clipping, the signals differentially propagate over the backplane (so-called Sum-backplane) to Sum-boards. There, they are summed up and passed to a discriminator. Finally, the digitalized signals propagate, again over the Sum-backplane, to a computer control board that also contains an FPGA, which counts the macrocells triggers and perform a global OR that generates the final trigger signal. This over-all trigger is connected to the MAGIC telescopes trigger coincidence logic (level 3) and from there to the readout.
  
\subsection{The Clip-boards and the Analogue Delay Modules}
The Clip-board is a 9U mixed digital/analogue board where the necessary corrections to the signal are performed (see Fig.\ref{ST_schematics_fig}, light orange shaded boxes). 
\begin{itemize}
	\item Delay: To correct individual PMTs timing differences and fibres delays. The delays range is about 6.5\,ns.
	\item Attenuation: To obtain perfect flat-fielding of all channel gains. Besides the differences on gains are normally small, we can apply attenuation from 0 to -32\,dB.
	\item Clipping: To avoid that the PMT's afterpulses generate fake triggers. Monte Carlo shows that this is the main source of noise \cite{bib:Dazzi}.
\end{itemize}

Because of ultra high channel density, it was necessary to incorporate the three above functions in so-called delay modules, of which 32 units are plugged onto the Clip-board. This is the most crucial component of the system. It contains a newly developed, adjustable analogue signal delay module.
\begin{figure}[h!]
	\centering
	\includegraphics[width=0.45\textwidth, angle=180]{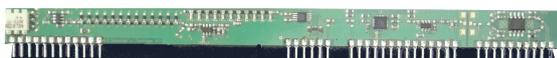}
	\caption{Analogue Delay Module.} 
	\label{delayline_fig}
\end{figure}
 
The analogue delay lines have been completely tested, showing high precision during timing, amplitude and clipping adjustment. The temperature dependency is negligible and the signal's integrity is acceptable.  The output pulse with maximum delay shows just a slight widening of 0.2\,ns compared to the input signal \cite{bib:Dennis}. At this time, mass production and quality control of the delay lines is taking place.

There are 18 Clip-boards for each Sum-Trigger (for each telescope). The main components of the Clip-board are these 32 programmable delay lines, and the CPLD (MAX-II, Altera) that allows the control of the board and the delay modules.
\begin{figure}[h!]
	\centering
	\includegraphics[width=0.25\textwidth]{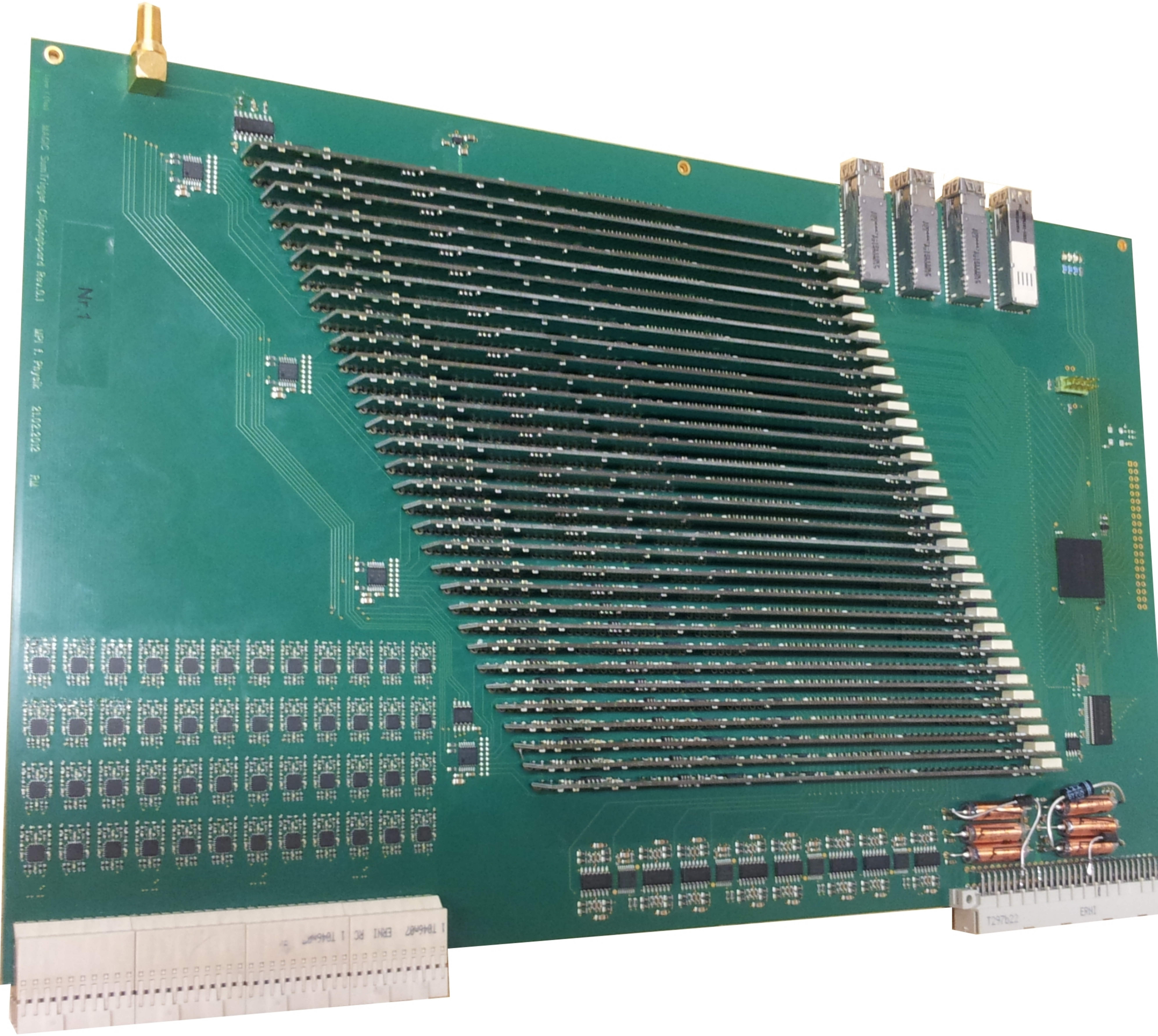}
	\caption{Picture of a Clip-board equipped with delay lines.} 
	\label{Clipboard_fig}
\end{figure}

%The Clip-board has passed standard electrical and power control checks and, also, fully equipped (with 32 delay modules) it passed a complete functionality test, where different delay, and amplitude settings were tested. 

% \begin{figure}[h!]
%   \centering
%   \includegraphics[width=0.45\textwidth]{icrc2013-0666-06.pdf}
%   \caption{ Clip-Board Functionality test. Each module represent different values of delay and amplitude: {\bf A} no delay, no attenuation; {\bf B} maximum delay, no attenuation;
% {\bf C} no delay, attenuation 8 dB; {\bf D } maximum delay, attenuation 8 dB;{\bf E} maximum delay, attenuation 16 dB.} 
%   \label{Clipboard_test_fig}
%  \end{figure}

%The test was complete successful.

The gain can be set with a precision below 5\% in a dynamic range from -15\,dB to +8\,dB. The delay (pulse position) can be adjusted with a resolution of 200\,ps. The measured RMS noise is $\le$ 0.05\,PhE and the crosstalk is $\le$ 1\%, thus fulfil by far the requirements.

\subsection{The Sum-backplane}
The Sum-backplane is a passive 10U printed circuit motherboard which connects the Clip-boards to the Sum-boards. From the electronic point of view, this is the most complex part of Sum-Trigger-II. The PCB layout has been developed considering that 997 fast differential analogue signals have to be routed to 55 macrocells, preserving the isochronism inside 50\,ps, a bandwidth higher than 650\,MHz and a low cross-talk ($\le$ 1\%) \cite{bib:Dazzi}. Both Sum-backplanes have already passed the electronic test and it fulfils the requirements.

\subsection{The Sum-boards}
\begin{figure}[h]
	\centering
	\includegraphics[width=0.36\textwidth]{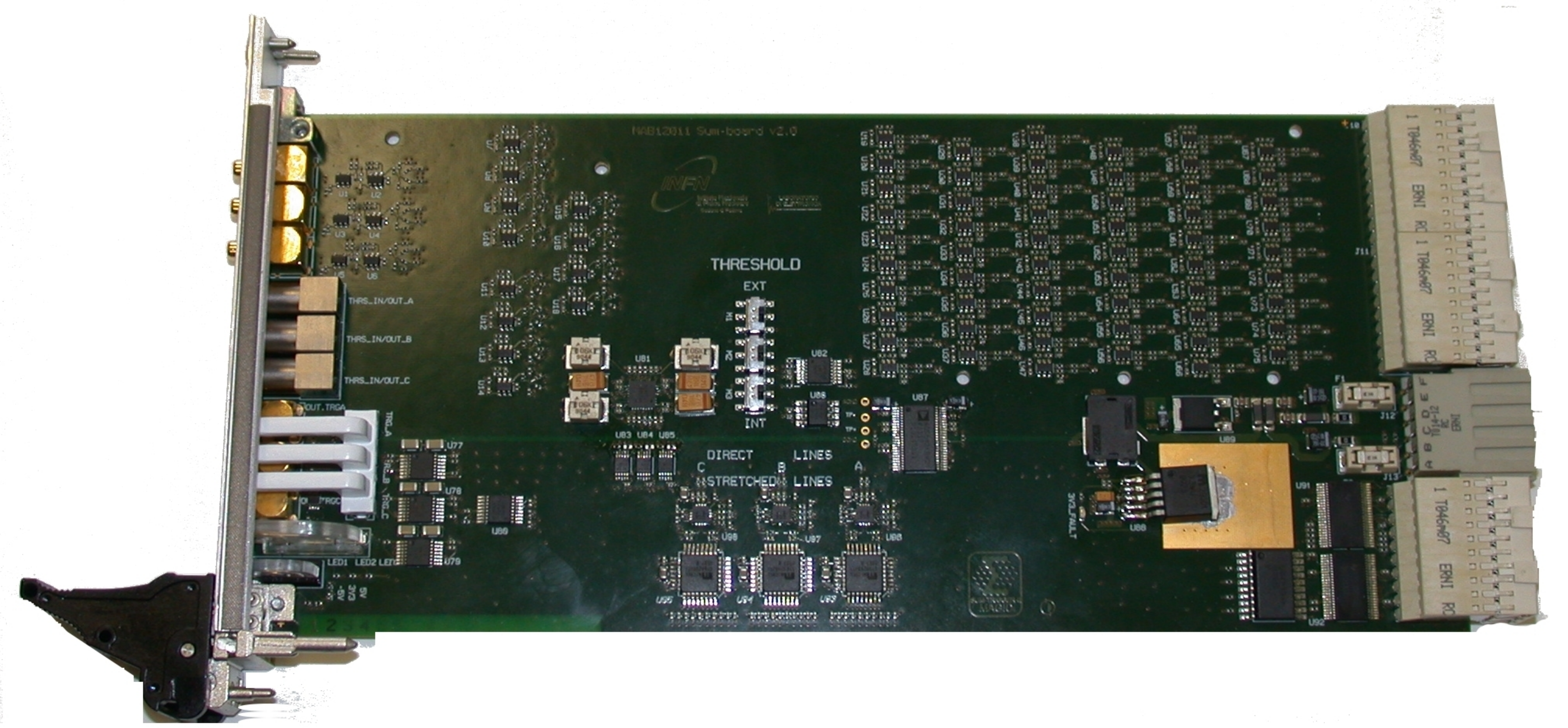}
	\caption{Sum-board.} 
	\label{Sumboard_fig}
\end{figure}

The Sum-board (Fig. \ref{ST_schematics_fig}, dark orange. Fig.\ref{Sumboard_fig}) is the main board for the final event selection. It can handle up to three macrocells, covering the whole mapping with only 19 units, still keeping the compactness of a small 3U card. It is a high-speed mixed analogue and digital system, which can be controlled automatically by the computer control board \cite{bib:Dazzi}.
 
The performances of the Sum-board meet the requirements. The bandwidth is around 550\,MHz and its flatness is inside 0.5\,dB up to 100\,MHz. The peak to peak electronics noise is quite low, similar to 0.6\,PhE. The cross-talk is less than 1\% up to 300\,MHz. The maximum skew in the analogue part (pile-up synchronization) is 80\,ps and its linearity is perfect up to 3.7\,V.  The comparator threshold has a resolution of 0.1\,PhE, a perfect linearity and stability.

%The Sum-Board passed all the electric checks (power and functionality).
%A specific functionality test has been performed, including the
%electrical continuity of the channels, the comparators functionality, 
%the trigger outputs, the correct pile-up of the pulses and the
%differential trigger outputs.
The digital trigger output signal goes to the Astro-board where the final decision will take place.

\subsection{The computer control Board}
The Astro-board (see Fig.\ref{ST_schematics_fig}, blue boxes) is a fully digital 9U board. It is the final stage of the Sum-Trigger-II electronics chain. It contains a Cyclone IV FPGA and a FOX-G20 Linux computer. The combination of a FPGA with the Linux computer is very powerful and also simple. The complete
Sum-Trigger control program communicating with the MAGIC central control is running on this computer.
\begin{figure}[h]
	\centering
	\includegraphics[width=0.35\textwidth]{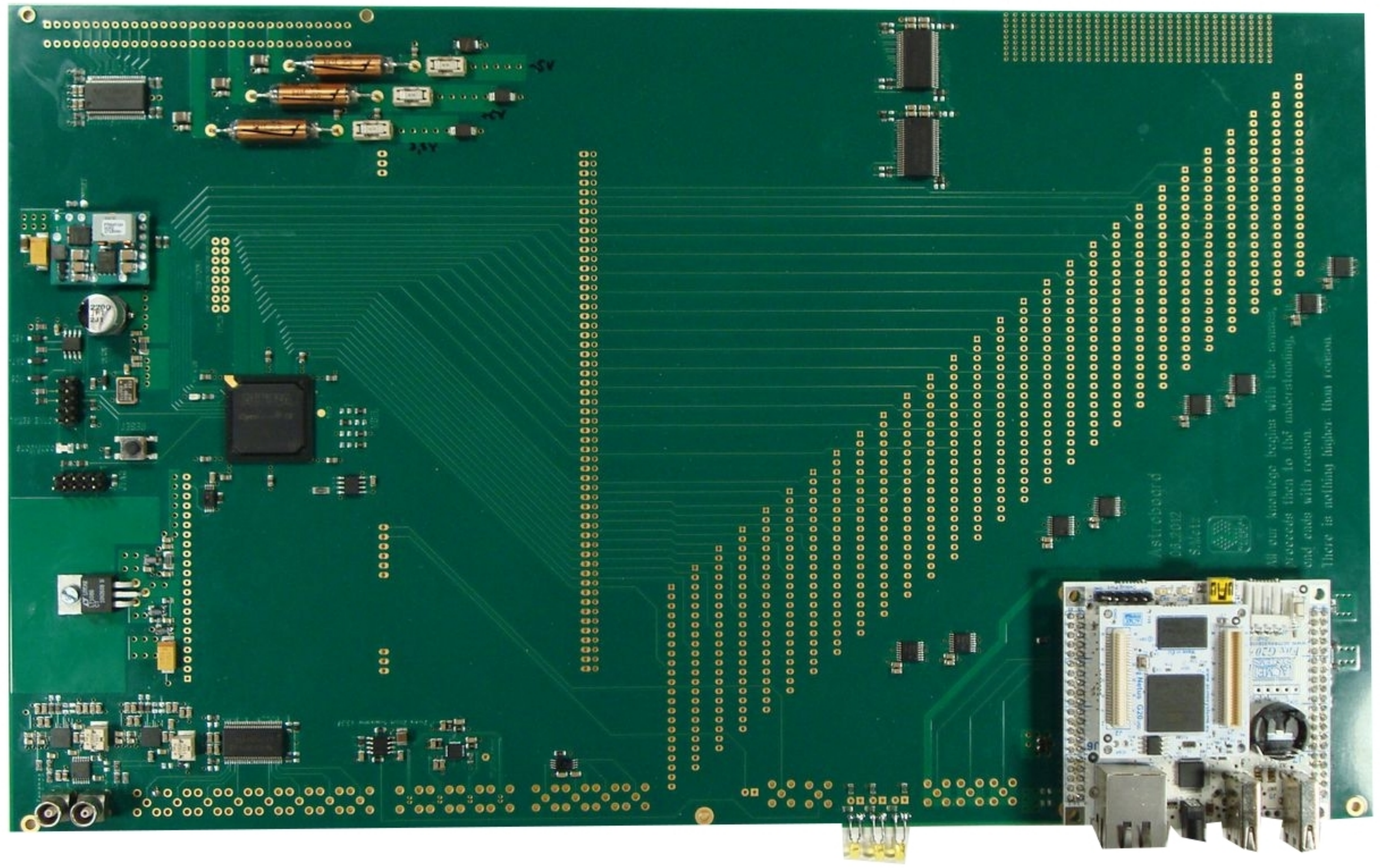}
	\caption{Astro-Board. The board contains an FPGA (Altera Cyclone IV) and an embedded Linux PC (FOX-G20), bottom right corner of the board.} 
	\label{ASTRO_fig}
\end{figure}

It take cares of:
\begin{itemize}
	\item The FPGA manages 55 macrocell triggers and adjusts the thresholds. At the same time, it merges these local triggers through a global OR, finally triggering the MAGIC-DAQ. 
	\item The trigger rate is kept quite stable and below the limits, adjusting continuously the discriminator thresholds of the Sum-boards. The correct parameters are computed by the embedded PC.
	\item Interface with Central Control Software: The Sum-Trigger gives reports and can be control by the Central Control of the telescopes thanks to this embedded computer. 
	\item Automatic calibration of delay, amplitude, and clipping can be perform since the PC can control the applied voltages on the Clip-board pins, so it
can fully control the delay lines. An automatic calibration procedure is under development.
\end{itemize}

% The automatic calibration procedure is a complicated algorithm that
% has to be applied only very rarely a few times per year. In the
% calibration procedure the delays of each channel are re-adjusted, the
% amplitudes are flatefielded and the clipping levels are calibrated in
% absolute units of PhE.

\subsection{Calibration process}
\begin{figure}[h!t]
	\centering
	\includegraphics[width=0.38\textwidth]{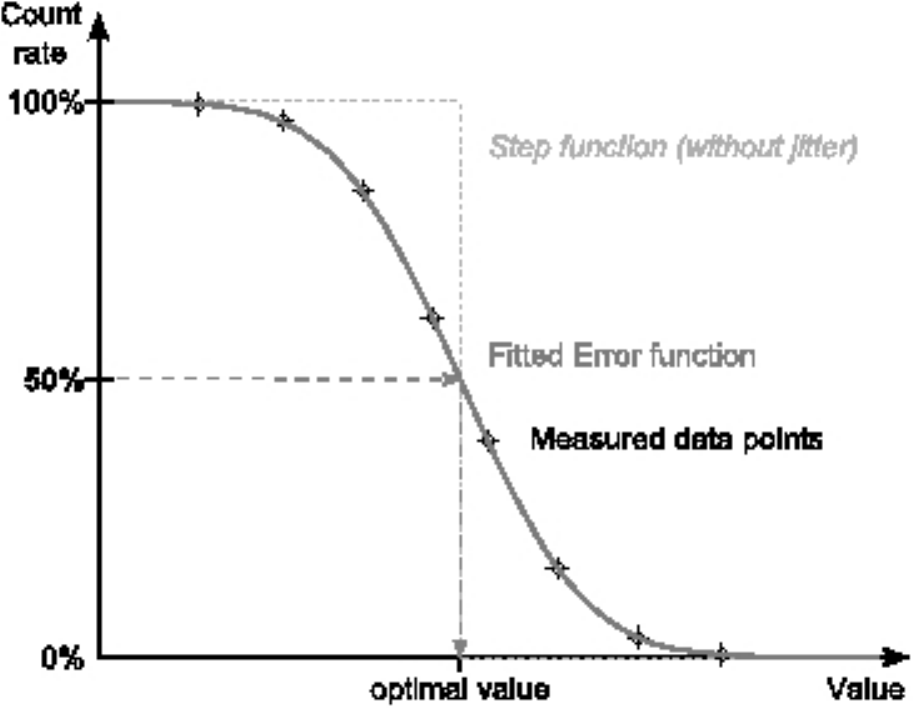}
	\caption{Principle of the measurement process. Here, ``value'' means \textit{gain} or \textit{delay}, depending on the property calibrated.}
	\label{fig_ratecountprinciple}
\end{figure}

Due to occasionally required tuning of the high-voltages applied to the PMTs, the amplitude and signal transit time can change for individual pixels and have to be re-adjusted on a regular basis for optimal trigger performance. Hence, Sum-Trigger-II includes a computer controlled automatic adjustment of amplitude and delay of the signals. In order to keep the complexity of the new circuits low, an innovative measuring technique based on the evaluation of a series of rate measurements is introduced, requiring only very few additional electronics. In particular, the discrete trigger output of the discriminator is used to measure amplitudes by counting the number of events that exceed the discriminator threshold within a certain time span. Similarly, the rates to determine the delay of each channel are derived from the output of a D-type flip-flop that is used to compare the arrival times of two pulses.\\
% To produce adequate reference signals, calibration LEDs, located in the center of the telescope reflector, emit light pulses at a well defined frequency and amplitude. Since the light pulses hit all camera pixels simultaneously with an equal intensity, they enable a relative timing and gain adjustment among all channels.\\
% The distinctive feature of the new measuring technique is to take advantage of intrinsic jitter in amplitude and arrival time of the signals, to efficiently derive the optimal settings for the trigger components. Relative amplitude variations are in the order of 10\%, and signal transit time jitters typically below 1 ns. These fluctuations primarily originate from the PMTs.\\
When performing the gain adjustment, the discriminator threshold is fixed to the target amplitude level, and the attenuation value is varied while counting the number of trigger signals from the discriminator. Likewise, the optimal delay is derived by tuning the delay line module. Here, the counter is incremented by the D-type flip-flop comparing the signal arrival time with the timing reference channel. The result is a series of rate measurements of the transition region from maximum to minimum number of trigger counts. Due to the time and amplitude jitter inherent in the signals, the rate scans show a cumulative distribution function, which is used to derive the optimal settings, being found at 50\% of the maximum rate (Fig.\ref{fig_ratecountprinciple}).
\section{Summary}
The analogue Sum-Trigger is a new trigger concept that has been invented, developed and tested on the single MAGIC telescope. It was designed to lower the trigger threshold down to 25\,GeV. The proof of principle was the detection of the Crab pulsar in 2008. \\
The Sum-Trigger-II has been designed for stereo observations. Thanks to the computer control it is able to compensate the signal delays, to equalize the gain and to stabilize the trigger rates such that observations at lowest thresholds inside the night sky regime are possible.

The Sum-Trigger-II has been tested and mass production of the individual elements and boards has started. The installation, on the MAGIC telescopes, is foreseen the end of this Summer (2013). This system will allow us to lower the trigger threshold of MAGIC and observe many interesting objects such as Pulsars, high redshift AGN and GRBs.

\vspace{4mm}
\footnotesize{{\bf Acknowledgement:}{ Technical support: M. Bettini, D. Corti, A. Dettlaff, D. Fink, M. Fras, P. Grundner, C. Knust, R. Maier, S. Metz, M. Nicoletto, O. Reimann, Ma. Reitmeier, Mi. Reitmeier, K. Schlammer, T. S. Tran, H. Wenninger, P. Zatti.}

\end{document}